# A Computer Aided Detection system for mammographic images implemented on a GRID infrastructure

U. Bottigli, P. Cerello, P. Delogu, M.E. Fantacci, F. Fauci, G. Forni, B. Golosio, P.L. Indovina, A. Lauria,
E. Lopez Torres, R. Magro, G.L. Masala, P. Oliva, R. Palmiero, G. Raso, A. Retico, A. Stefanini, S. Stumbo,
S. Tangaro

*Abstract*— The use of an automatic system for the analysis of mammographic images has proven to be very useful to radiologists in the investigation of breast cancer, especially in the framework of mammographic-screening programs. A breast neoplasia is often marked by the presence of microcalcification clusters and massive lesions in the mammogram: hence the need for tools able to recognize such lesions at an early stage.

In the framework of the GPCALMA (GRID Platform for Computer Assisted Library for MAmmography) project, the co-working of italian physicists and radiologists built a large distributed database of digitized mammographic images (about 5500 images corresponding to 1650 patients) and developed a CAD (Computer Aided Detection) system, able to make an automatic search of massive lesions and microcalcification clusters. The CAD is implemented in the GPCALMA integrated station, which can be used also for digitization, as archive and to perform statistical analyses. Some GPCALMA integrated stations have already been implemented and are currently on clinical trial in some italian hospitals.

The emerging GRID technology can been used to connect the GPCALMA integrated stations operating in different medical centers. The GRID approach will support an effective tele- and co-working between radiologists, cancer specialists and epidemiology experts by allowing remote image analysis and interactive online diagnosis.

## I. INTRODUCTION

BREAST cancer is reported as one of the first causes of woman mortality [1]. The early diagnosis of this pathology in asymptomatic women makes it possible the reduction of breast cancer mortality. In spite of a growing number of detected cancer, the death rate for breast neoplasia decreased in the last 10 years [2], thanks also to early diagnosis made possible by mammographic-screening programs [3]. In the framework of a screening program working in Italy, a mammographic examination is performed on 49 to 69 years-old women. Mammography is widely recognized as the only imaging modality that is useful for the early detection of the abnormalities indicating the presence of a breast cancer [4]. It is usually realized by screen-film modality but digital detectors are becoming widespread [5]-[7]. It has been estimated that screening-programs radiologists fail to detect up to approximately 25% breast cancers visible on retrospective reviews and that the percentage increases if minimal signs are considered [8]-[10]. Sensitivity (percentage of pathologic images correctly classified) and specificity (percentage of non pathologic images correctly classified) of this examination appreciably increase if two radiologists independently analyze the images [11]. Independent double reading is currently strongly recommended as it allows the reduction of the rate of false negative examinations by 5 to 15% [12,13].

The recent technological progress has led to the development of several Computer Aided Detection (CAD) systems [14]-[16], which could be successfully used as second readers in the framework of mammographic-screening programs.

We report the characteristics and the performances of the CAD software currently implemented in the GPCALMA integrated station. We also performed a comparison between the GPCALMA CAD and another commercial CAD system, the CADx SecondLook[TM], and we report their performances as second readers [17]. Finally we discuss the capability of the GRID in providing the necessary tools for allowing the tele- and co-working of experts from different GRID-enabled medical centers.

## II. THE CALMA PROJECT

The CALMA (Computer Assisted Library for MAmmography) project involved italian physicists and radiologists from several research institutes and medical centers between 1998 and 2001. The aim of the project was the building of a large database of annotated mammograms and the development of a CAD system for the automatic search of massive lesions and microcalcification clusters, which are the radiological signs of breast cancer. The experiment was very successful since the CALMA database is at present one of the largest in Europe and the CALMA CAD system has proven to be very useful especially as second reader in the framework of mammographic-screening programs.

III. THE CALMA DATABASE

The CALMA collaboration built a large database of digitized mammographic images. It currently consists of about 5500 images (corresponding to 1650 patients), acquired in the italian hospitals that joined the project.

The images (18 cm x 24 cm, digitised by a CCD linear scanner with an 85 μm pitch and 4096 grey levels) are fully characterised: pathological ones have a consistent description that includes radiological diagnosis and histological data, whereas non-pathological ones correspond to patients with a follow up of at least three years [16].

IV. THE CALMA CAD SOFTWARE

The CALMA CAD approach consists in the analysis of each mammogram available per patient. The presence of suspicious areas for massive lesions or microcalcification clusters in one or more mammograms results in a possible cancer diagnosis. The relatively large size of a mammogram brings to the need of reducing data input with no loss of information before proceeding to the classification, in order to perform an efficient diagnosis in a reasonable amount of time. Therefore the approach is a multi-level one. The first levels are demanded to reduce the amount of information without excluding ill regions (demand of sensitivity values close to 100%), whereas the last levels are requested to perform an exact classification in order to reduce the number of false positives as much as possible.

The analysis is organized into three general steps:
1. Data reduction: non-interesting regions of the mammogram are eliminated with the consequent reduction of the amount of data passed to the subsequent step. This is obtained either with a ROI (Region Of Interest) hunter consisting of a first set of algorithms which perform loose cuts and identify interesting regions or by down-sampling data points to be later on classified (in this case being helped by the relatively large size of the objects to be detected);
2. Feature extraction: relevant characteristics are extracted out of the selected regions;
3. Classification: the selected regions are classified.

The algorithms used can vary depending on the kind of analysis (microcalcification clusters or massive lesions, opacities or spiculated lesions, search or analysis, search of Regions Of Interest or suspected regions, classification on the basis of the level of suspiciousness, etc.). The selection and the optimization of the suitable algorithm to each kind and step of analysis are made following both general rules and empirical approaches. Spatial frequency analysis, statistical methods and Neural Networks (both supervised and unsupervised) are applied in order to solve a given problem, and the method showing the best performance is chosen. Hybrid approaches can consist of many different methods, such as loose cuts, data down-sampling, high- and low-pass filters, convolution filters, Fourier analysis, statistical selection of local and global maxima, Principal Component Analysis performed by self-organizing maps (Sanger's Neural Networks), Feed Forward Neural Networks with different architectures.

*A. Opacities and Spiculated Lesions*

Masses (Fig. 1) are rather large objects with very different shapes and show them up with a faint contrast slowly increasing with time.

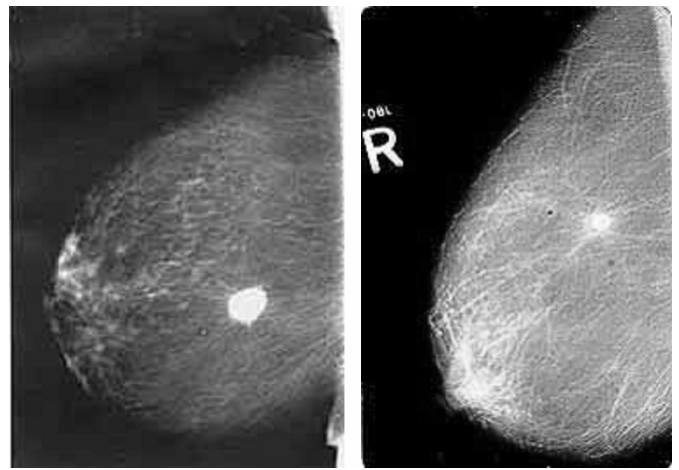

Fig. 1. Examples of massive lesions.

We developed algorithms for the recognition of opacities in general and specifically for spiculated lesions, which are characterized by a particular spike shape [18,19]. In order to detect opacities in a digital mammogram the interesting areas are selected by constructing a structure with concentric rings centered on local intensity maxima until the mean pixel value reaches a fixed threshold. The ROIs are thus identified; they consist in circles of radius R. For opacities, features are extracted by calculating the average intensity, variance and skewness (index of the asymmetry of the distribution) of the pixel value distributions in circles of radius 1/3 R, 2/3 R and R, respectively (Fig. 2). The features extracted by means of this procedure are used as input for a feed-forward neural network, which perform the final classification. The output neuron of this network is characterized by a threshold (i.e. a number in the 0 to 1 range) that represents the degree of suspiciousness of the corresponding ROI.

*B. Microcalcification Clusters*

A microcalcification is a rather small (0.1 to 1.0 mm in diameter) but very brilliant object. The presence of microcalcifications grouped in clusters (Fig. 3) may be a strong indication of breast cancer.

The microcalcification cluster analysis was made using the following approach:

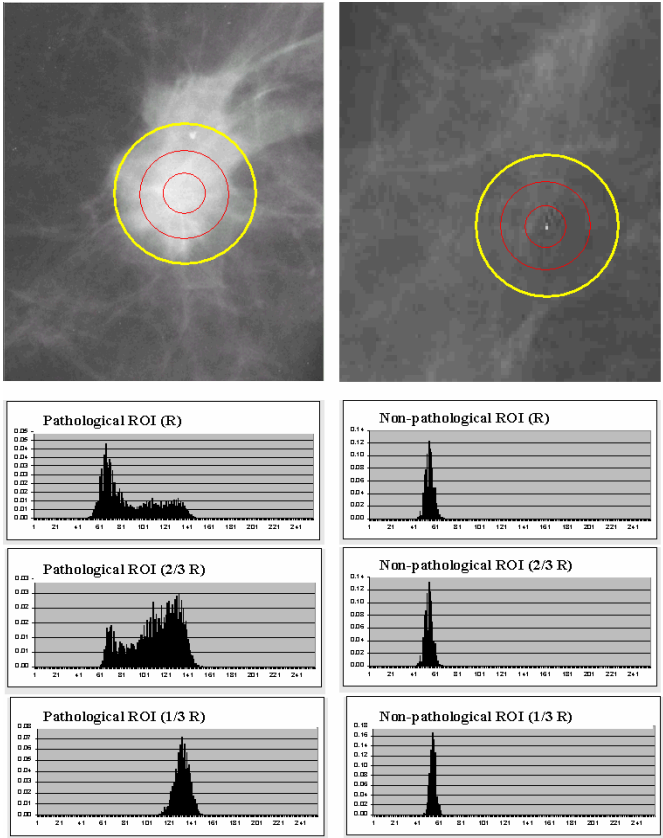

Fig. 2. Distributions of pathological and non-pathological ROIs of radius R, 2/3 R and 1/3 R.

1. The digital mammogram is divided into 60x60 pixels wide windows;
2. The windows are shrunk from 60x60 to 7x7 and are classified (with or without microcalcification clusters) using a FFNN with 49 input, 6 hidden, and 2 output neurons.

If the window under examination is classified as positive by the FFNN, the analysis follows in this way:

3. The windows are processed by a convolution filter in order to reduce the large structures;
4. A self-organizing map (a Sanger's neural network) analyzes each window and produces 8 principal components;
5. The principal components are used as input to a FFNN able to classify the windows matched to a threshold (the response of the output neuron of the neural network);
6. The windows are sorted by the threshold;
7. A maximum of three windows, whose thresholds exceed a given value, is memorized;
8. The overlapping windows are merged.

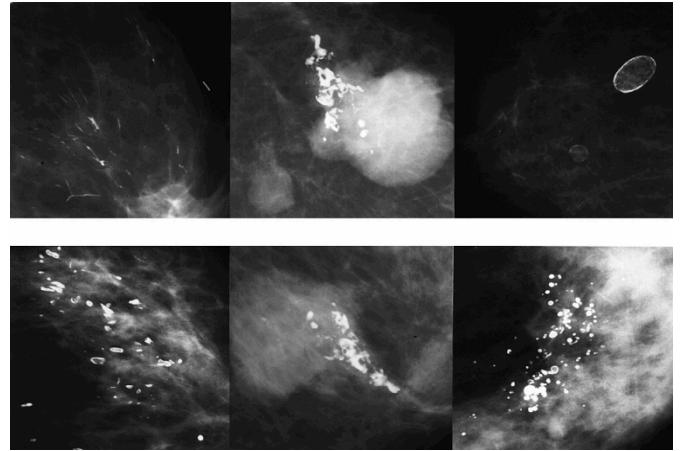

Fig. 3. Examples of microcalcification clusters.

## V. RESULTS

### A. Opacities and spiculated lesions

By using the CALMA database, the FFNN adopted in the procedure for the massive-lesion detection and previously described has been trained on a training set of 515 images (102 containing opacities and 413 without) and tested on a test set composed of 515 different images (again 102 containing opacities and 413 without).

The results of such a classification are shown in the ROC (Receiver Operating Characteristics) curve reported in Fig. 4, which shows sensitivity (true-positives fraction) versus 1-specificity (false-positives fraction) for different threshold values. The best results we obtained are 94% for sensitivity and 95% for specificity.

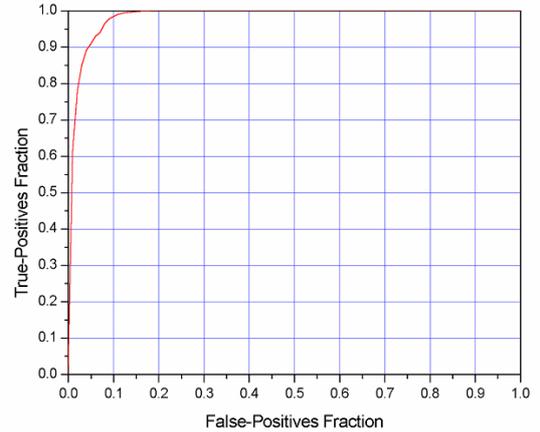

Fig. 4. ROC curve for massive lesions (realized with 515 mammograms, 102 containing opacities and/or spiculated lesions).

### B. Microcalcification clusters

A dataset of 676 images containing microcalcifications and 995 images without microcalcification clusters has been used in the previously described procedure for the detection of microcalcification clusters. In particular, in the training phase 865 images (370 with and 495 without microcalcification clusters) were used, whereas the test set consisted in 806 images (306 with and 500 without microcalcification clusters).

The results obtained in this test are summarized in the ROC curve reported in Fig. 5, which is realized by varying the threshold according to which the interesting windows are sorted. The best results we obtained are 92% for both sensitivity and specificity.

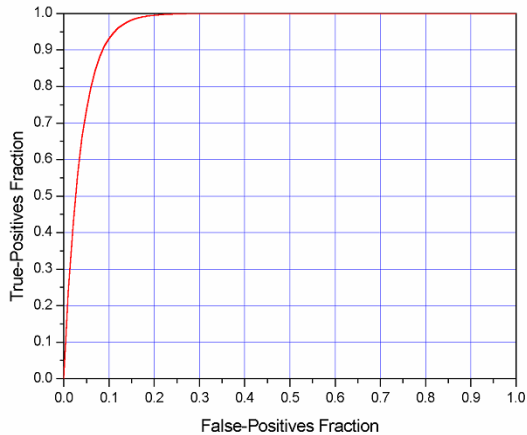

Fig. 5. ROC curve for microcalcification clusters (realized with 865 mammograms, 370 containing microcalcification clusters).

*C. Practical use of the CALMA CAD system*

The hardware requirements for the CALMA CAD station are very simple: a PC with SCSI bus connected to a planar scanner. As the size of information to be stored is large, big disks and a CD-ROM recorder are recommended. A high-resolution monitor helps the human radiological diagnosis. The station can process mammograms directly acquired by the scanner and/or images from files and allows human and/or automatic analysis of the digital mammograms. The human analysis allows a diagnosis of the breast lesions in terms of kind, localization on the image, average dimensions and, if present, histological type. The automatic procedure finds on the image the ROIs that are characterized by a chance (greater than the chosen threshold value) of containing a pathological sign. These data are stored as raw images, .pgm and .gif. Raw data are used in CAD analysis, the .pgm file is used to show on the monitor an 8 bit-depth resolution image, while the .gif file provides a quick preview of the images. For each image of the preview, it is possible to open the image, to enter or modify the diagnosis, the annotations and the ROI contours corresponding to the radiologist geometrical indication. There is also an interactive window that allows zoom, windowing, gray level and contrast selection. The operator can also start the CAD analysis by choosing in other interactive windows the threshold for ROI selection for microcalcifications or opacities. The station allows also for queries and statistical studies on the local database.

*D. Performance of the CAD systems as second readers*

We analyzed the performances of the CALMA CAD and of another CAD system, the CADx SecondLook[TM] as second readers [17]. SecondLook[TM] is a commercial system developed by CADx Medical Systems (Canada) and approved by FDA. Values reported for the sensitivity of this system vary from 88.7% for opacities to 98.2% for microcalcification clusters. Regarding specificity, values are reported in terms of false positive per image (FP/IMG): 0.28 for microcalcification clusters and 0.97 for opacities [20].

We performed the test on CADx SecondLook[TM] and CALMA CAD as second readers by using the following criteria: two radiologists with different degrees of experience made the diagnoses on a set of mammograms with and without the help of the two CAD systems. The results obtained show a dependence on the radiologist experience. As shown in Table I an increment of sensitivity up to 11.5% is reported when CADx is used, and up to 15.6% when CALMA CAD is used. With respect to the specificity (see Table II), a decrement up to 3.4% is reported when CADx is used and up to 3.3% when CALMA CAD is used. It is worth noting that no decrement in specificity has been reported for the radiologist A (the more expert one) when CALMA CAD is used as second reader.

TABLE I
SENSITIVITY

|   | Radiologist | Radiologist with CADx | Radiologist with CALMA CAD |
|---|---|---|---|
| A | 82.8% | 94.3% | 94.3% |
| B | 71.5% | 82.9% | 87.1% |

Sensitivity reported for the radiologists A and B (characterized by two different degree of experience), radiologists with CADx SecondLook[TM], and radiologists with CALMA CAD system.

TABLE II
SPECIFICITY

|   | Radiologist | Radiologist with CADx | Radiologist with CALMA CAD |
|---|---|---|---|
| A | 87.5% | 84.2% | 87.5% |
| B | 74.2% | 70.8% | 70.9% |

Specificity reported for the radiologists A and B (characterized by two different degree of experience), radiologists with CADx SecondLook[TM], and radiologists with CALMA CAD system.

## VI. THE GPCALMA PROJECT

The success of the CALMA project, but also the necessity of solving some relevant issues left open by this experiment, led italian physicists and radiologists to join a new collaboration, the GPCALMA (Grid Platform for Computer Assisted Library for MAmmography) project, which is supported by the italian INFN (Istituto Nazionale di Fisica Nucleare).

The emerging GRID technology would be extremely suitable for solving some CALMA open issues, such as:
1. The management of a rapidly increasing (and virtually unlimited) database;
2. The handling of the legal restrictions on access to data that reside in hospitals and are firewall-protected;
3. The request of making available to each user the whole intrinsically-distributed database;
4. The securing of file transfer of large images;

5. The request of enabling the tele-diagnosis in order to reduce significantly the delays in examining the acquired mammograms that can be associated to the screening programs (order of months).

The GRID philosophy based on "moving code rather than data" seems to be very suitable for images fitting approximately 10.5 Mbytes, that are stored locally in each hospital. At the same time the GRID approach would support an effective tele- and co-working between radiologists, cancer specialists and epidemiology experts by allowing remote image analysis and interactive online diagnosis.

The approach adopted by the GPCALMA project is based on ROOT for the application code, on PROOF for the parallel remote data analysis and on AliEn (the ALICE production environment developed by the ALICE collaboration at CERN) for the distributed data management. Thanks to the PROOF functionality, this approach will avoid data replication for all the images with a negative diagnosis (which are estimated to be about 95% of the sample), while it will allow a real-time diagnosis for the 5% of images with high cancer probability.

The CALMA CAD software has been installed on the GPCALMA station with a dedicated GUI (graphics user interface), which has a facility tool for the images visualization and elaboration, in order to support the medical diagnosis directly on a high-resolution screen. Radiologists can use the GPCALMA integrated station also for the digitization and storage of mammograms and to perform statistical analyses. The implemented CAD is also directly usable with digital mammography systems since it is compatible with the standard DICOM (Digital Imaging and Communications in Medicine). Some GPCALMA integrated stations have already been installed and are on clinical trial in some italian hospitals.

Very recently (March 2003) the first remote analysis between two different GPCALMA integrated stations has successfully been performed.

VII. CONCLUDING REMARKS

The GPCALMA collaboration has very recently achieved extremely encouraging results in the planned procedure of GRID enabling the CALMA experiment. By using a GPCALMA integrated station, a radiologist can store mammograms, run the CAD system for the detection of microcalcification clusters and massive lesions on every image stored in the shared database and perform statistical analysis. The basic functionality of the GRID connection that would allow the remote image analysis between different medical centers is currently available and tested. In order to go one step further, the GRID connection would be made available to all the hospitals joining the GPCALMA project.